\documentclass[journal]{IEEEtran}

\ifCLASSINFOpdf
\else
   \usepackage[dvips]{graphicx}
\fi
\usepackage{url}
\usepackage{amssymb,amsmath}

\usepackage{graphicx}

\begin{document}

\title{Online Automatic Speech Recognition with\\Listen, Attend and Spell Model}

\author{Roger Hsiao*, Dogan Can*, Tim Ng*, Ruchir Travadi and Arnab Ghoshal
\thanks{R. Hsiao, D. Can, T. Ng, R. Travadi and A. Ghoshal are with Apple Inc. (e-mail: rhsiao@apple.com; dogan\_can@apple.com; tim\_ng@apple.com; rtravadi@apple.com; aghoshal@apple.com). }
\thanks{*: these authors contributed equally to this work.}
}

\maketitle

\begin{abstract}
  The Listen, Attend and Spell (LAS) model and other attention-based automatic speech recognition (ASR) models have known limitations when operated in a fully online mode.
  In this paper, we analyze the online operation of LAS models to demonstrate that these limitations stem from the handling of silence regions and the reliability of online attention mechanism at the edge of input buffers.
  We propose a novel and simple technique that can achieve fully online recognition while meeting accuracy and latency targets.
  For the Mandarin dictation task, our proposed approach can achieve a character error rate in online operation that is within 4\% relative to an offline LAS model.
  The proposed online LAS model operates at 12\% lower latency relative to a conventional neural network hidden Markov model hybrid of comparable accuracy.
  We have validated the proposed method through a production scale deployment, which, to the best of our knowledge, is the first such deployment of a fully online LAS model. 
\end{abstract}

\begin{IEEEkeywords}
end-to-end ASR, online recognition
\end{IEEEkeywords}

\IEEEpeerreviewmaketitle

\section{Introduction}
\label{sect:intro}

\IEEEPARstart{T}{he} Listen, Attend, and Spell (LAS) model is a widely-used architecture for the so-called {\em end-to-end} automatic speech recognition (ASR)~\cite{chan2016}.
Compared to other end-to-end methods proposed in literature, LAS tends to be easier to train and can achieve higher accuracy.
It has been shown that LAS may outperform deep neural network based hidden Markov model (DNN-HMM)~\cite{bourlard1994, hinton2012} hybrids in large scale ASR tasks~\cite{chiu2018}.
However, online recognition with such a model remains a challenge due to the fact that the generation of outputs from the model is not synchronized
with the consumption of inputs by the model (referred to as asynchronous decoding subsequently).
In recent years, research progress has been made with alternatives like RNN transducer (RNN-T)~\cite{graves2012, he2019} that allows time synchronous decoding,
multi-model techniques~\cite{kim2017a, watanabe2017, moritz2019, tara2020, tara2020a, li2020} where a time-synchronous model like RNN-T or CTC provides timing information to the LAS decoder,
or specially designed loss functions that aim to penalize latency~\cite{inaguma2020, li2020}.
While these approaches are effective, they can involve more complicated training algorithms~\cite{li2019, inaguma2020}
or decoding procedures~\cite{tara2020, tara2020a, li2020}. In this paper, we aim to understand and address the fundamental issues
related to online recognition for LAS models and create a stand-alone LAS model that is capable of online recognition.

Online recognition LAS models requires three components: 1) streamable encoder, 2) online attention mechanism and 3) online asynchronous decoding.
The first two issues have been sufficiently addressed in the literature. It has been proposed to use a uni-directional encoder or
a latency controlled bi-directional encoder~\cite{fan2019}. For streamable attention mechanism, there are multiple proposals including,
but not limited to, monotonic attention~\cite{raffel2017}, monotonic chunkwise attention (MoChA)~\cite{mocha2018} and local attention~\cite{merboldt2019}.
However, addressing these two issues is insufficient for practical use of LAS in online recognition, without also addressing issues related to asynchronous decoding. 
This is a key difference compared to time synchronous decoding adopted by DNN-HMM hybrids, CTC, or RNN-T, where the time synchronous decoder creates
a monotonic map between audio features and output tokens as it consumes audio features.
We show in later sections that this asynchronous nature is a major obstacle to LAS and similar attention-based models to perform online recognition.
For such models, the ability to do online decoding is contingent upon the model learning when to terminate the decoding process. 

This paper has two main contributions.
First, we perform a thorough analysis of the main obstacles for online recognition with LAS models.
Second, we propose a novel, yet simple approach that would allow LAS model to perform online recognition.
The underlying concept is based on silence modeling and a buffering scheme that is applicable to any end-to-end model that uses asynchronous decoding,
especially the LAS model that uses an online attention mechanism, like MoChA and local attention.
We think this work could contribute to advancing the end-to-end technology, and could be useful to anyone who wants to deploy such technology.

\section{LAS model and online attention}
\label{sect:background}

LAS computes the posterior probability of a sentence given a sequence of acoustic features by
\begin{eqnarray}
  P(Y | X) & = & \prod_i^N P(y_i | X, y_{1, \dots, i-1})
\end{eqnarray}
where $X = (x_1, x_2, \dots, x_T)$ is a sequence of $T$ acoustic features; $Y = (y_1, y_2, \dots, y_N)$
is a sequence of $N$ output tokens. An output token could be a character or a subword and it also
includes beginning of sentence (BOS) and end of sentence (EOS) symbols.

The probability of each predicted token, $P(y_i | X, y_{1, \dots, i-1})$, is computed by an encoder
and a decoder. The encoder maps the acoustic features into a sequence of hidden features. The mapping
can be achieved through a pyramidal RNN~\cite{chan2016},
\begin{eqnarray}
  h_t^k, r_t^k & = & \text{RNN}_k([h_{2t}^{k-1}, h_{2t+1}^{k-1}], r_{t-1}^k)  \text{ if $k>1$} \nonumber \\
  h_t^1, r_t^1 & = & \text{RNN}_1(x_t, r_{t-1}^1)  \text{ if $k=1$}
\end{eqnarray}
where $h_t^k$ is the hidden features of $k$-th layer of the encoder; $r_t^k$ is the
RNN internal state at time $t$. This RNN takes the outputs from the previous layer by 
concatenating the neighboring features. As a result, the length of the output sequence of each layer is half
of the corresponding input sequence. After $K$ layers of encoder RNN,
the decoder would apply an attention mechanism to compute,
\begin{eqnarray}
  c_i & = & a(s_i, h_1^{T'}) \label{eqn:attention}
\end{eqnarray}
where $c_i$ is the context vector for $i$-th decoding step computed by the attention mechanism $a$;
$s_i$ is the internal state of the decoder;
$h_1^{T'}$ is the output hidden features from the pyramidal RNN where $T'=T/2^K$.
This context vector is then used by the decoder RNN to compute,
\begin{eqnarray}
  o_i, s_i = \text{RNN}(s_{i-1}, y_{i-1}, c_{i-1})
\end{eqnarray}
where $o_i$ is the output of the decoder RNN at $i$-th decoding step, and $s_i$ is the internal state of the
decoder RNN. Finally, we compute the posterior probability based on $o_i$ and $c_i$. The function $f$ can be as
simple as a softmax over an affine transform to project $o_i$ and/or $c_i$,
\begin{eqnarray}
  P(y_i | X, y_{1,\dots,i-1}) & = & f(o_i, c_i)
\end{eqnarray}

The attention mechanism described by equation~\ref{eqn:attention} is global if $a$ can access the entire utterance $h_1^{T'}$
at any decoding step. This is not compatible with online recognition for an obvious reason.
The monotonic attention of~\cite{raffel2017} limits the visible region of an utterance by allowing
one and only one frame of encoder output to be accessible by the decoder at each decoding step, and the chosen
encoder output at step $i+1$ has to be either the same chosen frame at step $i$, or some future frame. Therefore, it ensures the attention mechanism to be
monotonic. MoChA attention~\cite{mocha2018} can be considered as an extension of monotonic attention. Instead of only allowing one frame
to be accessible, MoChA attention allows the decoder to access a chunk at a time, and monotonicity is guaranteed at the chunk level.

 \section{Synchronous and Asynchronous Decoding}
\label{sect:problems}

We call a decoding process time synchronous if as it consumes audio features, it always generates a monotonic many-to-one mapping between the features and the output targets.
For example, the Viterbi decoder for DNN-HMM hybrid would produce a monotonic map between every frame of audio features and HMM states (usually tied context-dependent states or senones) that are the outputs of the DNN.
The same is true for CTC or RNN-T except the output targets for such models may be letters or word pieces and a special {\em blank} symbol. In such cases,
the decoding process stops once it consumes all audio features.

In contrast, the decoding process of LAS and other attention-based ASR models is asynchronous and it does not guarantee that any output is generated as inputs are consumed. Although the model creates the encoded representation of the inputs and may even calculate the attention weights, the decoding process may require the consumption of an arbitrary length of input before generating any output. Similarly, after having consumed a sufficient length of inputs, the decoding process may generate an arbitrary length of outputs.
In other words, given a segment of speech, an asynchronous decoder could produce none or an arbitrary number of output tokens. Such a condition is not possible with time synchronous decoders. As a result,
the stopping criterion of such a decoding process cannot be based on the exhaustion of input audio, but through learning when to self-terminate. 
For the LAS model, we have the BOS and EOS symbols because of this reason.

\begin{figure}
\centerline{\includegraphics[width=7.0cm]{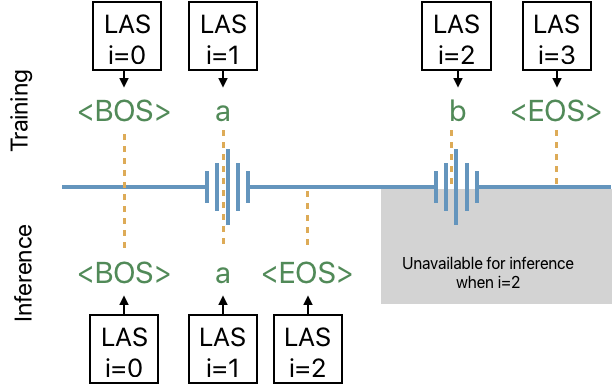}}
\caption{A hypothetical example showing the differences between training and inference.
         In this utterance, the reference is ``$<$BOS$>$ a b $<$EOS$>$'' and there is a long silence region between a and b.
         The upper part and the lower part show the training and inference process respectively. The black boxes show what the model
         is processing at each step (indexed by $i$) and the orange dotted line is the peak of the attention at different steps.
         This example shows during inference, the model could not jump to the next segment like training since the features
         have not arrived yet.
        }
\label{fig:training_and_inference}
\end{figure}

It is important to appreciate this key difference in order to understand why online recognition is difficult with asynchronous decoding. With LAS models, even though we may use monotonic or MoChA attention to guarantee monotonicity, without controlling the way the model learns to self-terminate it is not possible to guarantee complete online operation. Figure~\ref{fig:training_and_inference} illustrates the difficulty of online recognition by comparing the training and inference processes.
As shown in the example, this utterance consists of two speech segments, and there is a long silence region in between.
During training, the LAS model could jump from the first segment to the second one as the entire utterance is visible at training time.
This is true even for monotonic and MoChA attention since they compute expected attended regions for each output token during training.
As a result, the model may learn to skip attending to the silence segments and there is no constraint to prohibit this kind of behaviors.

The situation, however, is different during online recognition. As the acoustic features are being streamed,
the asynchronous decoder can only look at a portion of the utterance.
As long as the silence segment is long enough, the decoder would only
see a trailing silence after the first speech segment (as shown in the gray area in figure~\ref{fig:training_and_inference}).
At this point, the LAS model may terminate the decoding prematurely by emitting an EOS symbol, even though there are still
incoming features. In our experience, this leads to a large number of deletion errors, and it is the main obstacle for LAS model to perform online recognition.

\section{Silence Modeling for LAS}
\label{sect:sil_modeling}

Appreciating the issues related to the asynchronous nature of attention-based decoders allows us to view the existing methods in literature in the context of forcing a greater time-synchronosity of attention-based decoding.
This includes methods like triggered attention \cite{moritz2019}, as well as those that penalize the generation latency \cite{inaguma2020, li2020}.
We propose that the same effect can be achieved simply by modeling the silence regions and having the LAS model generate silence tokens.
We notice that silence modeling also improves the end of utterance latency without explicitly controlling for it (cf. Section \ref{sect:expr}).

We insert silence tokens to the references so that the training process would teach the model when, and how often, it should output a silence token.
This idea is similar to the explicit silence models used in HMM-based ASR systems~\cite{povey2011}. 
In HMM-based ASR, the silence model often has multiple states and loops so it could generate multiple silence frames. Our idea is similar
in principle. By adding multiple silence tokens, the LAS model would learn to output silence tokens repeatedly depending on the length
of the silence segment. The only difference compared to DNN-HMM is that we help the model learn this only through
the data instead of having an explicit structure in the model.

\begin{figure}
\centerline{\includegraphics[width=7.0cm]{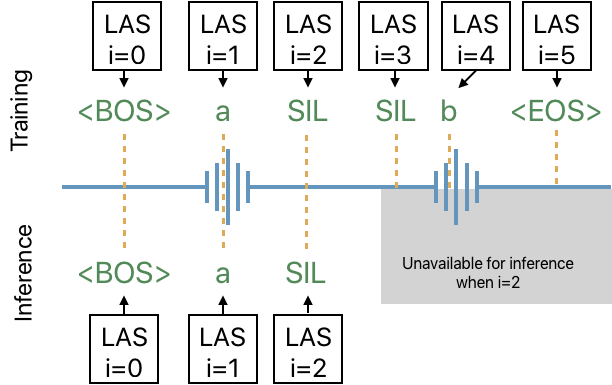}}
\caption{A hypothetical example with explicit silence modeling. Compared to figure~\ref{fig:training_and_inference},
         we allow the model to output silence tokens and insert silence labels to the reference. As a result, during online recognition,
         the model outputs a silence token instead of an EOS token when it sees a trailing silence segment.
        }
\label{fig:training_and_inference_with_sil}
\end{figure}

Figure~\ref{fig:training_and_inference_with_sil} is an example to explain the idea. In this example, we insert multiple silence tokens to the reference.
Each silence token corresponds to a fixed number of silence frames, and each silence segment could have one or more silence tokens depending on its length.
This number of frames for each silence label is called duration of silence in this paper.
After training, the model would learn to output silence tokens when it sees a silence segment. Therefore, during online recognition, the decoder could choose to output
a silence token instead of an EOS symbol in figure~\ref{fig:training_and_inference_with_sil}.
To insert silence segments to the references, we could use a hybrid DNN-HMM system to perform forced alignment. Based on the alignment information,
we could choose to insert a silence token for every N number of silence frames. This preprocessing procedure would require a hybrid DNN-HMM model.
In practice, this is not a problem since one could build a useful hybrid DNN-HMM system
with small amount of data~\cite{cui2013} and the pronunciation lexicon could be replaced by a graphemic dictionary~\cite{kanthak2002, killer2003}.

While adding this explicit silence model would help online recognition, the model could still output an EOS symbol prematurely.
In figure~\ref{fig:training_and_inference_with_sil}, at decoding step $i=2$, the model could output EOS instead of silence.
To help reducing this confusion, we propose a buffering scheme to help online recognition.
In this buffering scheme, the audio data is streamed to the decoder
in batches of $T$ ms long. For each batch, we designate a restricted buffer at the end of the batch.
During decoding, if the peak of the attention distribution of any attention head falls within this restricted buffer,
this suggests the attention may benefit from more data, and so
the decoding for this batch is voided and backtracked. The decoder would then wait for more audio data, unless it is the end of a utterance,
which would perform decoding until the model emits EOS.

The purpose of this buffering scheme is to ensure there is enough audio data, so the decoder would not output a token
when the corresponding audio is truncated. This idea can be extended so that the buffer size depends on the previous output token.
For example, if the decoder is processing a silence segment, we might use a larger buffer so the model is less likely to emit EOS prematurely.
Therefore, we could have different buffer sizes for regular and silence tokens.
When the encoder has processed all audio features, the buffering scheme would be disabled so the decoder can access all the encoder outputs
and finish the decoding process. Generally speaking, smaller buffer means smaller latency, however, the buffer size does not imply a lower bound of the latency,
since the last batch may be smaller.

 \section{Experimental Results}
\label{sect:expr}

We test our proposed approach on our proprietary Mandarin voice assistant and dictation tasks. 
We use a research training set consisting of 5500 hours of transcribed audio, sampled from both use cases.
We evaluate the models on two test sets | one for the voice assistant use case and another for the dictation use case | each with 10 hours of audio data. 

For the audio features, we extract 40 dimensional filter-bank coefficients with a standard 25ms window and 10ms frame shift.
The LAS model has three layers of pyramidal LSTM with a reduction factor of two for each layer. Each encoder LSTM has
1200 hidden units and 600 dimensional recurrent projection. The decoder uses MoChA attention with a chunk size of three and the dimension of the
context vector is 600. The decoder has three LSTM layers, each with 800 hidden units and 300 dimensional recurrent projection.
The output layer is 7975 dimensional and each output corresponds to
a Chinese character, Latin character, digit, punctuation, a special token like BOS and EOS, or a silence token.
The entire model has around 64 million parameters.

To train the model, we use block momentum algorithm~\cite{chen2016} to optimize the model with cross entropy loss.
The training uses scheduled sampling~\cite{bengio2015} with a probability of 0.2, 
and it also uses spectral augmentation~\cite{park2019} and label smoothing~\cite{szegedy2016} with a smoothing factor of 0.2.
During decoding, we use a decoding beam of eight, unless otherwise specified.
The audio data is streamed to the decoder in 320ms batches. Each batch of audio data is processed by the model's encoder and
the encoder output is then appended to the buffer. As mentioned in section~\ref{sect:sil_modeling},
decoding can only happen when the buffer is larger than a predefined minimum buffer size.

\begin{figure}
\centerline{\includegraphics[width=7.5cm]{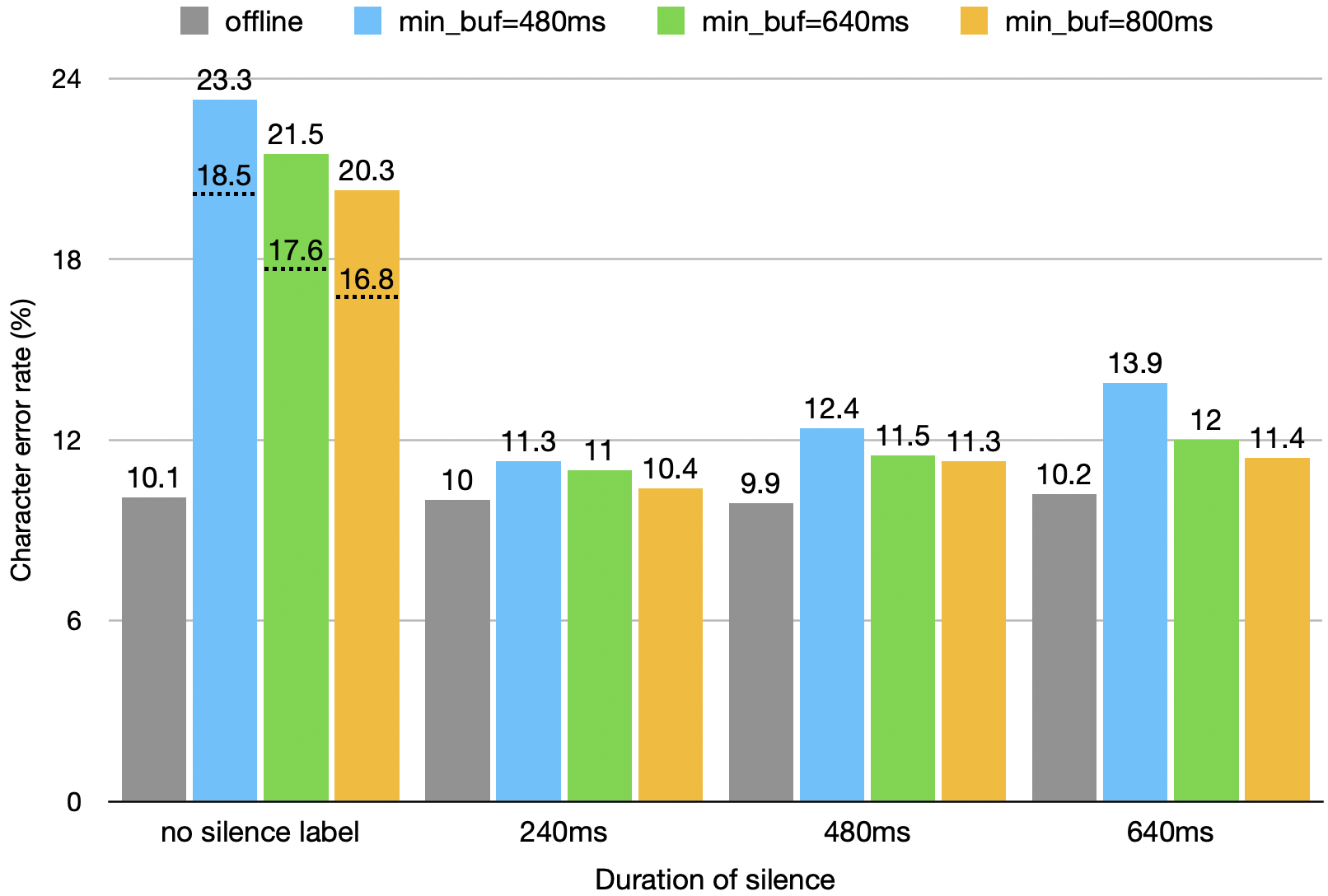}}
\caption{An experiment on the dictation evaluation set to evaluate how duration of silence affects online recognition. The black dotted lines are the error rates of restarting
         the decoder when the baseline model emits an end of sentence token prematurely. The beam is set to eight for this experiment.
        }
\label{fig:sil_dur_and_streaming}
\end{figure}

Figure~\ref{fig:sil_dur_and_streaming} studies how different durations of silence,
which is the length of the silence segment for each silence label, interacts with different minimum buffer sizes. 
The results give us some interesting insights. First, without an explicit silence model (``no silence label''),
although the LAS model could achieve 10.1\% character error rate (CER) on our dictation test set for offline decoding,
the accuracy is greatly degraded for online recognition regardless of the minimum buffer size (all over 20.0\% CER).
The reason for much higher CER is due to high deletion rate (12-15\% deletion rate)
and these deletions are caused by early stopping as discussed in section~\ref{sect:problems}. For this baseline model, we also try to restart
the decoder whenever the model emits an EOS token when there is still unconsumed audio features. While this technique alleviates part of the deletion problem,
it still has significant degradation compared to the offline accuracy.
This shows that adopting online attention mechanisms and uni-directional encoder alone are not enough for LAS model to perform online decoding.
Second, we find that with silence modeling, the accuracy gap between online and offline decoding is greatly reduced. For 240ms duration of silence, 
the CER is 10.4\%-11.3\% depending on the minimum buffer size, which is close to the offline result of 10.1\% CER.
Third, while bigger buffer would generally improve accuracy, it might impact latency,
so smaller duration of silence is favorable as we can use a smaller buffer.

\begin{table}[htbp]
\begin{center}
\caption{CER(\%) of different silence buffering sizes. The buffer for other tokens is 480ms and silence duration is 240ms.}
\vspace{-1mm}
\begin{tabular}{|c|c|c|c|c|} \hline
Test set & Offline  & sil buf=480ms & sil buf=640ms & sil buf=800ms \\
         \hline \hline
         Assistant  & 9.3 & 11.9 & 10.9 & 9.8 \\ \hline
         Dictation  & 10.1 & 11.3 & 11.1 & 10.5 \\ \hline
\end{tabular}
\label{tbl:sil_mem}
\end{center}
\end{table}

In table~\ref{tbl:sil_mem}, we investigate the effect of having different buffer sizes for speech and silence segments. As discussed in section~\ref{sect:sil_modeling},
the purpose is to use a larger buffer for silence segments to reduce the risk of generating an EOS symbol prematurely.
While the regular minimum buffer is set to 480ms, we use a larger buffer when the last output is silence.
This adjustment improves the CER from 11.3\% to 10.5\% for dictation and from 11.9\% to 9.8\% for assistant test sets.

\begin{figure}
\centerline{\includegraphics[width=\columnwidth]{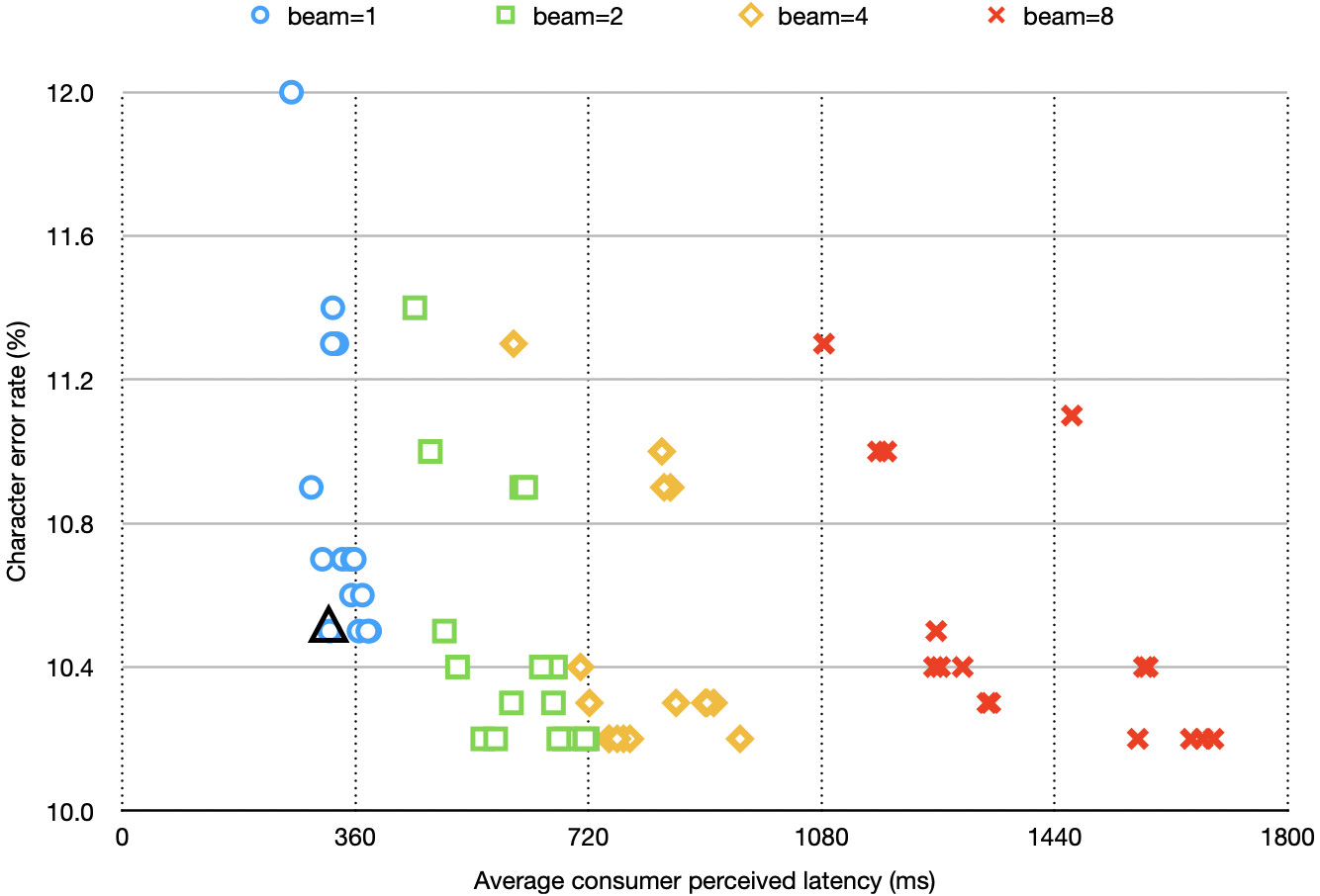}}
  \caption{An experiment on the dictation evaluation set to compare consumer perceived latency (CPL) vs character error rate (CER). For each beam,
           we perform a grid search on the minimum buffer size and silence buffer size.
           This measurement is done on an Intel Xeon E5-2640 2.4GHz server.
           The black triangle is the chosen operating point and its CER is 10.5\% and the average CPL is 320ms.
        }
\label{fig:cpl_vs_wer}
\end{figure}

Figure~\ref{fig:cpl_vs_wer} studies how latency is related to accuracy given our findings so far. In this experiment, we enumerate the decoder setting
including beam size, minimum buffer size and silence buffer size. Then for each setting, we measure the CER and latency. In this work, latency is the time between the last word spoken
and the time when the decoder no longer changes the hypothesis presented to the user. This definition of latency aims to capture perceived latency for the user and hence, we call it
consumer perceived latency (CPL). In this paper, we report average CPL.  The data point marked in a black triangle is the chosen operating point, where the beam is one,
and both minimum buffer and silence buffer are set to 960ms. The resulting CER is 10.5\% and the average CPL is 320ms.
Compared to the offline baseline, the difference in CER is only 0.4\% absolute or 4.0\% relative.

\section{Conclusions}
\label{sect:conclusions}

In this paper, we explain the challenges of online recognition for LAS models. We show that online encoder and attention mechanism alone are not enough for
online recognition, and the model can suffer from early stopping.  However, with our proposed silence modeling and buffering scheme,
we show that LAS model is capable for online recognition. In our dictation evaluation set, the online LAS model has 10.5\% CER and 320ms average CPL,
which has only 0.4\% absolute or 4.0\% relative difference compared to the offline baseline.

\section{Acknowledgments}
\label{sect:acknowledgments}

We would like to thank Ossama Abdelhamid, John Bridle, Pawel Swietojanski, Russ Webb and Manhung Siu for their support and useful discussions. We also want to thank
Ron Huang and Xinwei Li for their initial experiments on MoChA and spectral augmentation respectively.

\bibliographystyle{IEEEbib}
\bibliography{bib/abbrev,bib/e2e,bib/asr}

\end{document}